\documentstyle[11pt,mrs2001,epsfig]{article}
\begin{document}

 \hspace{-3cm}
\begin{center}
{\tt

``Where's the Matter?

Tracing Dark and Bright Matter with the New Generation of Large
Scale Surveys''

Marseille, June 2001 , Eds. Treyer \& Tresse, 2001, Frontier
Group.}
\end{center}

\hspace{3cm}

\title{The XMM-LSS Survey:\\
Mapping hot, luminous, obscured and dark material out to $z \sim
1-2$}

\author{Marguerite PIERRE $^1$}
\affil{$^1$ CEA/DSM/DAPNIA/SAp Saclay, F-91191 Gif Sur Yvette
~~~~~mpierre@cea.fr}

\begin{abstract}
We review the unique cosmological implications of the XMM-LSS
survey in association with its multi-$\lambda$ follow-up: (1)
Large Scale Structures traced by X-ray clusters and AGNs, optical
galaxies, weak lensing as well as Sunyaev-Zel'dovich effect; (2)
location of IR star forming galaxies and IR (obscured) AGNs within
the cosmic web. \\ The XMM/MegaCam/VIRMOS/SIRTF data base will
provide the first comprehensive study of structure formation -
from hundreds of Mpc to galaxy scale - in close connection with
environmental processes.

\end{abstract}

\section{Introduction}

The origin and evolution of the large-scale distribution of matter
is a major cosmological issue. Although the universe appears
homogeneous and isotropic on the largest scales, local galaxy
surveys have revealed the existence of foam-like structure;
galaxies are confined within sheets and filaments surrounding
large ``voids" with scales of 100 h$^{-1}$ Mpc \cite{lan}. Galaxy
clusters are usually located at the intersections of these sheets
and filaments. Within the current theoretical hierarchical
paradigm, structure originated in the very early universe and is
observed directly at an early time via the cosmic microwave
background (CMB) radiation. Density fluctuations were subsequently
amplified by gravity and followed by the process of galaxy
formation to produce the structure observed in the present epoch.
The present-day ``cosmic web" is therefore generated by the
details of several key cosmological processes; the origin of
structure, the nature and amount of dark matter, the nature of
galaxy formation and the specific values of cosmological
parameters. Observations of large-scale structure (LSS) therefore
form a key element in our global understanding of the Universe.

The traditional method employed to study LSS is to map the galaxy
distribution, either by covering very large coherent scales, e.g.
the Sloan and 2dF surveys (several thousands of deg$^{2}$ out to
$z_{max}\sim 0.1$ and 0.25 respectively), or probing evolution to
significant depths like deep pencil beams and the forthcoming
VIRMOS survey ($z_{max}\sim 1$ over a few deg$^{2}$). While
providing strong constraints on models for structure formation,
this approach is extremely data-intensive (from 150 000  galaxy
spectra for VIRMOS to  $10^{6}$ for Sloan) and the interpretation
depends both on global cosmological parameter combinations and on
the details of galaxy formation, the latter still poorly defined.
Another approach is to use QSOs: since they constitute  the most
luminous objects in the Universe, they can be  observed   up to
very high redshift. However, data interpretation  is even harder,
as the link between galactic nuclear activity and initial mass
fluctuations is currently not understood.

Alternatively, clusters of galaxies - the most massive entities in
the Universe - offer considerable advantages both because they can
provide complete samples of objects over a very large volume of
space and because they are in some respects ``simpler" to
understand. The matter halos of clusters are easily traced   by
their X-ray  emission (luminosity, size) while the theory
describing their  formation (biasing) and evolution from the
initial fluctuations is well tested by N-body simulations. Such a
level of understanding does not exist for galaxy and even less for
QSO formation.  Studies of cluster LSS and cluster abundances are
powerful tools to constrain cosmological parameter values,
independently of CMB and SN studies, as they do not rely on the
same processes. In particular, they can break the degeneracy
between the shape of the power spectrum and the matter density in
addition to constraining the validity of fundamental assumptions
of the standard paradigm, for example, that LSS grows only by
gravity.

Currently, the normalizations of the power spectra of structures
traced by different objects are significantly different. Can all
these results be consistent with a single Gaussian power spectrum
for mass fluctuations, how do normalizations change with scales,
what is the physical meaning of the normalization constants?

{\bf We have designed a comprehensive project with the
observational goal of mapping the distribution of matter over a
large volume out to redshifts $\sim 1-2$ using three complementary
techniques: (1) X-ray observations to locate galaxy clusters and
QSOs; (2) optical observations to obtain the galaxy distribution
and to locate dark matter through a weak lensing analysis; and (3)
Sunyaev-Zel'dovich effect observations to measure the distribution
of diffuse extragalactic hot gas.}

\section{The XMM Large Scale Structure survey}

\underline{Clusters of galaxies} are thought to be the largest
virialized entities in the universe. In the cosmic network
picture, cluster formation occurs by matter accretion along the
filaments so that the dynamical state of clusters is a permanent
competition between accretion and relaxation processes, which act
at rates that depend strongly on the embedding cosmology. Within
clusters, the masses of individual galaxies are negligible, since
dark matter accounts for $\sim 80$\% of the mass and most of the
rest is gas at a temperature of several $\times 10^7 \ \rm K$
which emits in the X-ray band. Averaged properties of X-ray
clusters can be much more easily simulated and modelled than the
galaxies they contain. For  instance, a correlation exists between
mass and X-ray luminosity for nearby clusters \cite{rei}, which
will certainly be investigated at high redshifts with up-coming
deep XMM observations. X-ray clusters are also easy to find. An
X-ray medium sensitivity survey at high galactic latitude
essentially shows two types of objects: galaxy clusters (extended)
and QSOs (pointlike).  As it  virtually eliminates projection
effects, it is a vital complement to optical/NIR searches for
(high redshift) clusters. Combined with optical spectroscopy, the
whole provides the only comprehensive mapping of the large scale
distribution of the deepest potential wells in the Universe and of
AGNs. Exploiting the unrivalled sensitivity of XMM, the XMM-LSS
will be $\sim 1000$ times deeper than the REFLEX survey
\cite{boh}: with a sensitivity of $\sim 3 \times 10^{-15} \, {\rm
ergs} \, {\rm s}^{-1} {\rm cm}^{-2}$ for point sources in the
[0.5-2] keV (95{\%} completeness level , \cite{val}), we expect
$\sim 300$ sources per deg$^{2}$. Out of them, approximately 15-20
clusters, 200 AGNs, with the remainder being stars and nearby
galaxies.
\par

Spanning  $8 \times 8 \ \rm deg^2$,  the XMM-LSS survey will probe
co--moving transverse scales of $\sim 320$ and 506 h$^{-1}$ Mpc at
redshifts $z=1$ and $z=2$ respectively. Approximately 900
clusters/groups will be identified out to $z=1$ ($\Lambda$CDM,
Fig. 1) and will enable the first investigation of the evolution
of the cluster correlation function in two redshift bins [0-0.5]
\& [0.5-1]. The survey design is constrained by the requirement to
obtain 15{\%} precision in the estimated correlation length in
each redshift bin. The low redshift bin will sample the cluster
population over a smaller region  than REFLEX but down to much
lower-mass groups. The high-$z$ bin will provide the first
measurement of the cluster correlation function at these
redshifts, and will enable a direct comparison with the low-$z$
correlation function of massive clusters by REFLEX.  The expected
accuracy for the cosmological parameters, $\Omega_{M}$,
$\sigma_{8}$, and $\Gamma$ will be 15{\%}, 10{\%} and 35{\%}
respectively. The XMM-LSS will also systematically explore the
existence of massive clusters to redshifts $z\sim2$ -- a new,
exciting territory. Though a handful of clusters are known around
$z \sim 1.2$,  properties and the state of equilibrium remain in
the realm of speculation beyond $z>1.5$.
 The constraints on
cosmology expected from the entire XMM-LSS cluster study are
summarized in Fig. 1 and discussed in detail by \cite{ref}.

More than 200 \underline{Active Galactic Nuclei/QSOs} are expected
per square degree at the proposed XMM-LSS sensitivity, with half
of these at $z < 1$ \cite{leh} and, in total, a space density 6
times higher than 2dF QSOs. This will provide the first complete
deep sample of X-ray QSOs over a large coherent volume of the
Universe and, consequently, an essential basis for comparison with
optical clustering studies (e.g. 2dF, Sloan) on scales ranging
from a few hundreds of pc to hundreds of Mpc. Together with a high
S/N correlation function, we shall study the location of QSOs
within the filament network defined by the cluster/group
population as a function of the QSOs' X-ray and optical
properties. This is crucial  for our understanding of galactic
nuclear activity  in terms of peculiar initial density
perturbations, environment conditions, and local galaxy
interaction rate as well as to address the debated QSO lensing
issue.

The XMM-LSS will also provide first clues about the existence and
the properties and space distribution of \underline{super clusters
of galaxies} in the distant universe.

Further important issues in cluster and QSO multi-wavelength
evolution will be also addressed by the survey. Although in
principle this could be achieved by XMM serendipitous pointings,
the XMM-LSS Legacy data set will readily possess the advantage of
uniform X-ray coverage and complete high quality optical imaging
and spectroscopy. The scientific issues are reviewed in detail by
\cite{pie00} and \cite{pie01}.

\section{Optical, radio and infra-red follow-up}
 \underline{The imaging of the $8 \times 8 \ \rm deg^2$ XMM-LSS
area} is the priority target of the Canada-France-Hawaii Legacy
Survey\footnote{http://cdsweb.u-strasbg.fr:2001/Instruments/Imaging/Megacam/MSWG/forum.html}.
MegaCam, the one degree field image built by CEA to be installed
at the new CFHT prime focus, will come into operation by mid-2002.
It will provide the deep high quality optical multi-color imaging
counterpart of the X-ray sources (u*=25.5, g'=26.8, r'=26.0,
i'=25.3, z'=24.3) at a rate of 15 deg$^{2}$/yr in at least three
colours. In particular,  an optical cluster catalogue is currently
under construction employing the CFH12k (then MegaCam) data using
both spatial clustering analysis and multi-color matched filter
techniques in addition to providing photometric redshift
estimates. Moreover, the MegaCam data will form the basis of a
weak lensing analysis\footnote{
http://www.iap.fr/LaboEtActivites/ThemesRecherche/Lentilles/LentillesTop.html},
whose cosmological constraints will be compared to that provided
by the X-ray data on the same region. This will be the first,
coherent study on such scales. We have also R and z' imaging from
CTIO. Data pipelines and processing  have been developed by the
TERAPIX\footnote{http://terapix.iap.fr} consortium; this will
provide object catalogues and astrometric positions for the entire
surveyed region. In addition, deep NIR VLT imaging (J, H, K ) of
$1<z<2$ cluster candidates found in in the XMM-LSS will be
performed in order to confirm their reality prior to spectroscopy.
\par
\underline{The standard spectroscopic follow-up} will perform
redshift measurements for all identified $0<z<1$ X-ray clusters in
Multi-Object-Spectroscopy mode, mainly using NTT/EMMI and
VLT/VIMOS. We plan to take 1 mask per cluster, randomly sampling
the AGN population at the same time, the underlying filamentary
galaxy distribution connecting clusters, radio sources from our
VLA survey as well as, possibly, a representative sample of the
SWIRE sources. This mapping around  $0<z<1$ clusters will have an
enormous scientific potential for studies of galaxy environments
and bias. We shall subsequently undertake programmes of advanced
spectroscopy (TNG, Las Campanas, CFHT, AAT, WHT, 2$\times$Gemini,
Magellan, LBT, VLT) that will focus on individual objects, and
include high resolution spectroscopy, the measurement of cluster
velocity dispersions, QSO absorption line surveys, as well as NIR
spectroscopy of our $z>1$ cluster candidates.
\par

 \underline{In the radio waveband} the complete survey region is
being mapped using the VLA at 74MHz and 325MHz. Radio coverage is
not only particularly relevant for tracing merger events triggered
by structure formation, but also a useful indicator of galactic
nuclear or star--formation activity.
\par
 \underline{Sunyaev-Zel'dovich} observations
(S-Z) are also planned. Clusters in the XMM-LSS field will be
targets of the prototype OCRA (One-Centimeter Radiometer Array)
instrument from 2002. The full XMM-LSS field will be mapped by the
complete OCRA, and will be an early target of the Array for
Microwave Background Anisotropy (AMiBA) after 2004 \cite{lia}.
This will enable a statistical analysis of the physics of the ICM
as a function of redshift. In the long term these observations
will also provide invaluable information on the low density
structures such as cluster outskirts and their connections to the
cosmic filaments. These measurements are complementary to the
X-ray and weak lensing data regarding the masses of clusters and
the structure of the hot gas they contain. The three data sets
together should provide  a direct and independent check of the
extragalactic distance scale.

 \underline{In the infrared}, the SWIRE\footnote{
http://www.ipac.caltech.edu/SWIRE} SIRTF  Legacy Programme will
cover 10  deg$^{2}$   of the XMM-LSS in 7 wavebands from 4 to 160
mm. The estimated IR source numbers in this area are around
20000/900/250 and 700/50/500 for starbursts/spiral-irregular/AGN
in the $0<z<1$ and $1<z<2$ redshift intervals respectively. This
represents a unique X-ray/IR combination in depth and scales to be
probed. The coordinated SWIRE/XMM-LSS observations will clarify an
important aspect of environmental studies: how star formation in
cluster galaxies depends on the distance to the cluster centre, on
the strength of the gravitational potential, and on the density of
the ICM (as inferred from the X-ray data). In this respect the
XMM-LSS represents the optimum SWIRE field, where galaxy
environment, deep NIR imaging and optical spectroscopic properties
will be the main parameters in modelling the MIR/FIR activity.
Here also, the location of IR AGNs within the cosmic web will help
establish their nature. The FIR/X/optical/radio association will
also provide unique insights into the physics of heavily obscured
objects as well as the first coherent study of biasing mechanisms
as a function of scale and cosmic time for X-ray hot (XMM), dark
(weak lensing), luminous galactic (optical/NIR) and obscured
(SWIRE) material.

In summary, the XMM-LSS multi-$\lambda$ data set will offer the
first evolving view of structure formation from Mpc to galaxy
scales. Its comprehensive approach constitutes a decisive new step
in the synergy between space and ground-based observatory
resources and therefore a building block of the forthcoming
Virtual Observatory.

\section{Survey layout}

\underline{The overall survey design} is a compromise between the
cosmic scale to be probed, the volume that must be surveyed to
detect enough clusters to measure cosmological parameters at a
significant new level of accuracy (see Sec.~2, Fig. 1), and the
total XMM exposure time.

\underline{The {\sc LSS Survey} location} is equatorial and has
been carefully chosen based on X/optical/IR visibility criteria:
it is a square area centered around $\rm \alpha = 2^h20^m$,
$\delta=-5^\circ$ (at $b=-58^\circ$, with neutral hydrogen column
$2 \times 10^{20} < N_H/{\rm cm^{-2}} < 5 \times 10^{20}$). This
area surrounds two deep XMM surveys based on guaranteed time: the
XMM\_SSC/Subaru Deep Survey (80~ks exposures in $1 \ \rm deg^2$)
and the XMM Medium Deep survey (XMDS; 20~ks exposures in $2 \ \rm
deg^2$), the latter being a collaboration between several Co-Is of
the present proposal; XMM: Li\`ege(OM), Milan-IFCTR(EPIC),
Saclay(SSC); MegaCam(Saclay, IAP); VIRMOS(France/Italy). The area
overlap will greatly assist in the study of selection effects.

\underline{The first release of the XMM-LSS} reduced data set will
occur by mid-2003 and will be subsequently updated on a yearly
basis as the X-ray coverage  and associated optical follow-up
proceed.

\section{More information}

is available at the XMM-LSS web page:\\
 {\bf
http://vela.astro.ulg.ac.be/themes/spatial/xmm/LSS/index\_e.html}

\vspace{3cm}

\begin{figure*}[h]
\centerline{\psfig{file=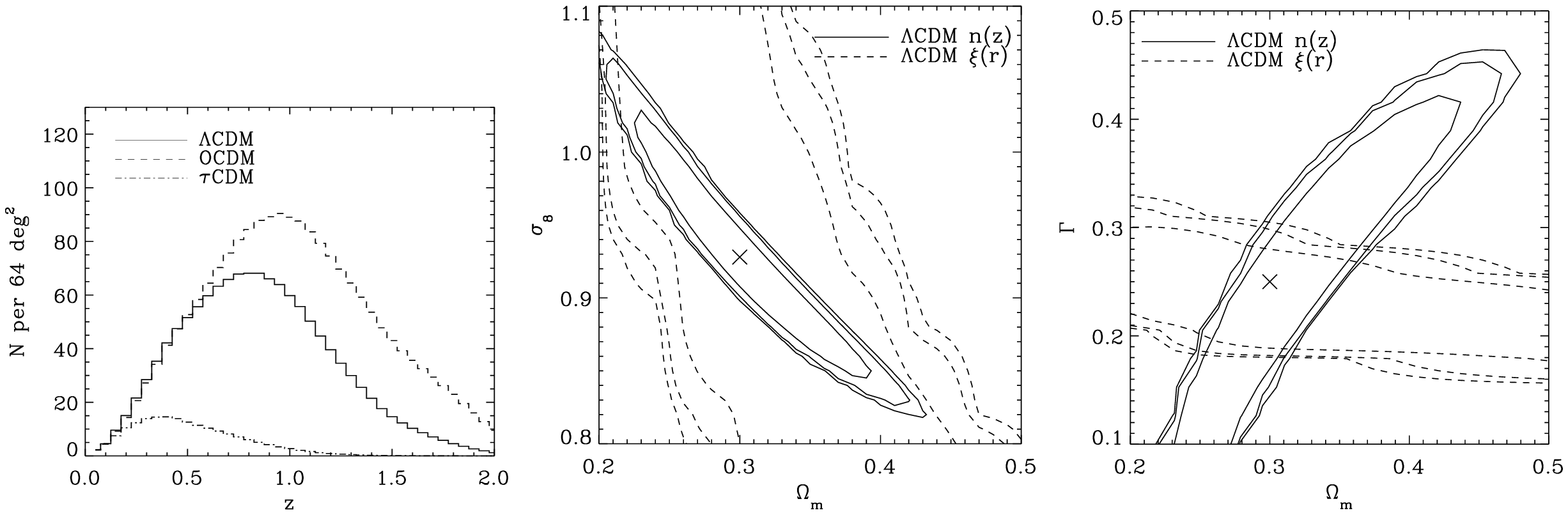,width=7cm,angle=90}}
\caption[]{{\bf Cosmological constraints from the XMM-LSS}(from
\cite{ref}).

{\bf Left:} The predicted XMM-LSS cluster redshift distribution
generated by various cosmological models. The selection function
for the XMM-LSS, generated by extensive image simulations, was
employed to construct each model.  Note that,  current favoured
values have been assumed here for $\Lambda$CDM, but the predicted
number of objects strongly depends on the assumed value of the
cosmological parameters (Fig. 7 of \cite{ref}); {\bf Centre:}
Constraints upon the cosmological parameters $\Omega_{m}$ and
$\sigma_{8}$ (the amplitude of mass fluctuations on 8 h$^{-1}$ Mpc
scale) for a $\Lambda$CDM universe obtained from XMM-LSS cluster
counts (solid lines) and correlation function (dashed lines). In
each case, the 68\%, 90\% and 95\% confidence level contours are
shown along with the assumed model (cross). Cluster abundance data
provides strong constraints upon the $\Omega_{m}-\sigma_{8}$
combination. {\bf Right:} Constraints upon the cosmological
parameters $\Gamma$ (the shape of the power spectrum) and
$\Omega_{m}$ for a $\Lambda$CDM universe (symbols as defined in
{\bf Centre}). The correlation function is a powerful tool to
constrain the shape of the initial spectrum. These calculations
have been performed assuming that only the redshifts over the
[$0<z<1] \times [$64 deg$^{2}$] volume are available.

In addition, identification of a ``Coma--type'' cluster within the
XMM-LSS over the redshift range $1.5<z<2$ has a probability of
$\sim 6.5 \times 10^{-7}$ in the current $\Lambda$CDM scenario.
Therefore, any such discoveries in the  survey would put the
currently favoured cosmological model in great observational
difficulty. }

\end{figure*}

\vfill
\end{document}